\documentclass[twocolumn,showpacs,preprintnumbers,amsmath,amssymb]{revtex4-1}
\usepackage{epsfig,amsopn}
\usepackage{graphicx}
\usepackage{sidecap}
\usepackage{amsmath,amssymb}
\usepackage{amsthm}
\usepackage{enumerate}
\usepackage{bbold}
\newcommand\bea{\begin{eqnarray}}
\newcommand\eea{\end{eqnarray}}
\newcommand\beq{\begin{equation}}
\newcommand\eeq{\end{equation}}

\newcommand{\bib}{\bibitem}

\def\nn{\nonumber}
\def\f{\frac}

\def\om{\omega}

\def\si{\sigma}
\def\Do{\partial}
\def\De{\Delta}

\def\la{\langle}
\def\ra{\rangle}

\def\ua{\uparrow}
\def\da{\downarrow}

\def\til{\tilde}

\begin{document}
\title{\bf{Magnetic field induced  Fabry-P\'erot resonances in helical edge 
states  }}
\author{ Abhiram Soori$^1$,
 Sourin Das$^2$ and Sumathi Rao$^3$}
\affiliation{$^1$ Centre for High Energy Physics, Indian Institute of Science,
Bangalore 560012, India.\\
$^2$ Department of Physics and Astrophysics, University of Delhi,
Delhi - 110 007, India.\\
$^3$ Harish-Chandra Research Institute, Chhatnag Road, 
Jhusi, Allahabad 211 019, India.}
\begin{abstract}
We  study electronic transport across a helical edge state exposed to 
a uniform magnetic ({$\vec B$}) field over a finite length. We show 
that this system exhibits Fabry-P\'erot type resonances in electronic 
transport.  The intrinsic spin anisotropy of the helical edge states 
allows us to tune these resonances by  changing the direction of the
{$\vec B$} field while keeping its magnitude constant. This is in sharp 
contrast to  the case of non-helical one dimensional electron gases with 
a parabolic dispersion, where similar resonances do appear in individual
spin channels ($\uparrow$  and $\downarrow$) separately which, however,
cannot be tuned by  merely changing the direction of the {$\vec B$} field.
These resonances provide a unique way to probe the helical nature of the 
theory.
\end{abstract}
\pacs{73.23.-b, 73.63.Nm, 71.10.Pm}
\maketitle
{\it{Introduction :-}} Edge states of a new class of insulators called 
topological insulators  form an interesting class of 1-D systems called
helical edge states ({HES})\cite{manyref}. The central feature of these
edge states is the fact that  the direction of propagation of the 
quasi-particles is directly correlated with their  spin projection, i.e.,
counter-propagating particles have opposite spin projections. Various  
aspects of this state have been studied\cite{vari}. Experimental evidence
 has been found for the existence of these edge states in a multi-terminal
 Hall bar
 setup\cite{konig}.

The fact that the spin of the electrons can be controlled and manipulated
 by manipulating its momentum due to spin-momentum locking has generated 
 great interest in the possible application of HES to the field of  
spintronics\cite{spint}. Besides electrical control, it is 
interesting to study the possibility of controlling  the spin of electrons
 on such edge states using magnetic  fields. This could be of great 
interest from  the point of view of application to spintronics devices. 
Since spin rotation symmetry is strongly broken in such edge states,  
a strong anisotropic response to an applied magnetic field is expected, 
which, in turn could be exploited for various applications. Hence one of 
the more intriguing features of the HES as opposed to the usual one 
dimensional electron gas with parabolic dispersion lies in its  response 
to magnetic fields.  Naively, the introduction of a magnetic field breaks
 the time-reversal symmetry which is central to the topological stability
 of the quantum Hall insulators which hosts these HES on its boundary. 
However,  it is still of interest to study the response of the HES to 
small magnetic fields which do not significantly disturb the  bulk 
stability of the topological insulators  but do influence the edge states
 in a nontrivial way. 

From the viewpoint of device-applications,  the study of  tunneling 
across barriers or back scattering from tunnel barriers implanted on the 
edge states is of central interest as they can act as experimentally 
tunable quantum resistors which are essential elements of any quantum 
circuitry. A simple way to produce controlled backscattering in mesoscopic
devices is to apply local  gate voltages. But for HES such techniques are 
not effective because all electrical barriers are rendered  transparent 
due to Klein tunneling of massless Dirac electrons. Further, protection 
of HES against inelastic backscattering due to electron-phonon coupling 
has also been reported~\cite{BDRB}. Hence, an alternate way is needed
 to produce back-scattering, and therefore,   magnetic barriers which 
do give rise to back-scattering are of vital importance in the context 
of HES. One of the more interesting aspects which has not yet been explored
 in  this context is the fact  that the effective size of the barrier 
depends 
also  on the direction of the applied {$\vec B$}-field  and not 
just its magnitude. 
Transmission through magnetic barriers has been studied
in  the case of chiral modes in carbon nanotubes~\cite{parafilo} and 
surface states of 3-D topological insulators~\cite{mondal2010}; however, 
a similar  study for HES is lacking and such a study is the central focus 
of this letter. Also, recently there have been theoretical studies 
involving  spin-polarised STM as a direct probe for testing the theoretical
prediction of the helical nature of the surface states, both in the context
of 2-D and 3-D topological insulators~\cite{Das2011}. Experimentally, only very
recently\cite{brune2012} the spin polarization at the edge was studied purely
by electrical means; hence such proposals are 
of crucial importance. This letter provides yet another way to probe the 
helical nature of HES via the study of spin anisotropy driven 
Fabry-P\'erot type resonances.

In this letter, we study the effect of a magnetic field patch on 
edge-state transport. We find that any {$\vec B$}-field with a non-zero 
component in the plane perpendicular to the spin quantization axis of 
left and right moving  states opens up a gap in the spectrum. Further, 
the spin of the left and right moving states gets twisted. In other words, 
even in  the presence of the {$\vec B$}-field,  the direction of motion 
remains tied with its spin; however, unlike in free HES, the 
counter-propagating states no longer  have spins anti-parallel to each 
other. We show that these states in the {$\vec B$}-field patch induce 
Fabry-P\'erot  type resonances in transport across the patch which are 
tunable purely by tuning the direction of the {$\vec B$}-field. The 
resonances, hence, provide a direct way to quantify the degree of spin 
anisotropy in these systems. \\
{\it{The model :-}}
We consider a situation where an infinitely extended helical edge state is 
exposed to a uniform {$\vec B$}-field over a finite length $L$. As long as 
the edge is smooth,  the free edge can be described by a Hamiltonian given 
below~\cite{manyref} 
\beq H_0 = -i\hbar v_F \int dx~\psi ^{\dag}(x)\si_{_Z} \Do_x
\psi(x)~, \label{H0} \eeq 
where $\psi=[\psi_{R\ua} ~\psi_{L\da}]^T$ and the $R/L$ index corresponds to 
right and left movers. Note that here we have used  $x$  to parametrize the 
coordinate along the 1-D edge state. $X,Y,Z$ are chosen to describe the spin 
in such a way that $Z$ points along the spin quantization axis of right-moving 
eigenstates in absence of $\vec B$ field. We choose $X$ to be along the 
component of applied $\vec B$ in the plane perpendicular to $Z$. 
\begin{figure}[htb]
\includegraphics[width=0.4\textwidth,height=0.13\textwidth]{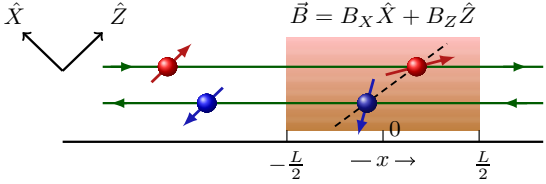}
\caption{Schematic  of the model. Shaded region $|x|<L/2$ is the region of 
constant magnetic field $\vec B$. Spin orientations of the left (right)-moving 
modes are tilted by $B_X$  in clock (anti-clock)-wise direction by the same
 angle.} \label{schematic}
\end{figure}
Since the spin and the direction of propagation are correlated in a HES, we 
will henceforth drop the $L/R$ -indices in $\psi_{L/R~\si}$. We now introduce 
a uniform magnetic field in  the region $|x|<L/2$ (sketched in 
Fig.~\ref{schematic}). The Zeeman coupling of the magnetic field to the 
intrinsic spin of the electrons can be modeled by
\bea
H_B &=& g \mu_B\int dx ~\De_{L} (x) \vec S(x) \cdot \vec B~, 
\label{HB}
\eea
where  $\De_{L} (x) =[\Theta(x+L/2)-\Theta(x-L/2)]$,  $\vec S(x) = 
1/2 (\psi^{\dag}(x) \, \vec \sigma\, \psi(x))$ is the spin operator, 
$g$ is the $g$-factor of the electron and $\mu_B$ is the Bohr magneton.  
The corresponding Heisenberg equations of motion for the fields $\psi(x,t)$  
is given by
\bea
 i \hbar \Do_t\psi &= & [-i\hbar v_F\si_{_Z} \Do_x +  
{\frac{g}{2}} \mu_B  \De_L(x)  {\vec\sigma} \cdot {\vec B}] 
\psi~~~\label{eq.mo}
\eea
By rescaling the energy $E\rightarrow E/W$, and by defining the dimensionless 
variables $x \rightarrow \bar x = xW/\hbar v_F$, $B\rightarrow \bar B = 
\mu_B B/W$ and $k\rightarrow \bar k = \hbar v_F k/W$,  where $W$ is
the bulk gap, we can rewrite this equation in terms of dimensionless variables
 as
\beq 
\bar E\psi = [-i \sigma_{_Z}  \partial_{\bar x}  +\frac{g}{2}\De_L(\bar x) 
{\vec\sigma} \cdot {\vec{\bar{B}}}] \psi .
\eeq
We will henceforth drop the bar's.

{\it{Magnetic field twists helicity :-}}
The Hamiltonian given in  Eq.~\eqref{H0} describing free edge state electrons 
is time-reversal-symmetric.  But the Zeeman-coupling term $H_B$  in 
Eq.~\eqref{HB} breaks time-reversal symmetry and  opens up a gap in the 
spectrum. This happens in the region where the magnetic field is nonzero, 
i.e., $|x|<L/2$. In regions of zero magnetic field we know that up-spin 
(pointing along $+Z$ direction) moves right and down-spin (pointing along 
$-Z$ direction) moves left. But in the region with non-zero magnetic field, 
this is no longer true.  However, as we shall see, the direction of motion
is still correlated with the spin orientation for propagating states 
above and below the gap. To look for the eigenstates of the Hamiltonian
$H_0+H_B$ in the region $|x| < L/2$,  we start with a plane wave solution 
$\psi_{k}e^{i (k x - \om t)}$ where $\psi_{k}$ is the Fourier
transform of the spinor field $\psi(x)$. Substituting the above form into 
Eq.~\eqref{eq.mo}, we find the dispersion relation to be- 
$E = \pm  \sqrt{  \til k^2+b_X^2}~,$ where $b_X = gB_X/2 $, $b_{_Z} = gB_{_Z}/2$ 
and $\til k = k+b_{_Z}$. Thus, the spectrum is gapped for nonzero $B_X$ 
with a gap given by $b_X$ and the spectrum breaks into two bands. 
\begin{SCfigure}
\centering
\includegraphics[scale=0.47]{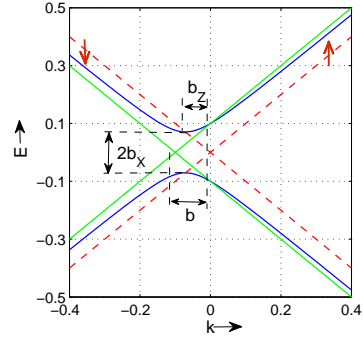}
\caption{Blue~(dark-solid)  and green~(light-solid) curves stand for 
schematic dispersions when the $\vec B$-field is at angles 
$\phi = \pi/4$ and $\phi=0$ with the positive $Z$-axis respectively. 
$b=|\vec b|=0.1$ in both cases.
Red~(dashed) curve corresponds to the zero field case.  } \label{disp}
\end{SCfigure}

The magnetic field $\vec B$ can be in general at an angle $\phi$ with 
$Z$-axis i.e., $\vec b = b(\sin \phi \hat X + \cos \phi \hat Z)$. 
The dispersion relation for the three different cases corresponding to 
$\vec b =0$ and directions $\phi=0$  and $\phi=\pi/4$ with non-zero $b$,
is shown in Fig.~\ref{disp}. Note  that in the presence of a finite 
${\vec  B}$ field, the zero momentum state, i.e., $k=0$ states are split 
exactly by $2b_X$ as expected for spin half particles with
 zero momentum. 

At a given energy $E$ the there are two eigenstates -
right- and left-moving $\psi_{E,R/L}$ with momentum $k=k_R$ and 
$k=-k_L$ respectively given by - 
\bea 
\psi_{E,R} & =& \f{1}{N}\Big[  E+ \til k_0~~~~~b_X \Big]^T;~
\psi_{E,L}~=~\si_{_X}\psi_{E,R}~~\nn \\
&{\rm and~}& k_{R/L} ~=~ [\mp b_{_Z} + sign(E) {\sqrt{ E^2-b_{_X}^2}}],
\label{LR-ud} 
\eea
where  $N$ is the normalization and $\til  k_0 = ( k_R+ k_L)/2$.
Unlike the free HES, here each state $\psi_E$ has its spin pointing 
in a distinct direction and the left and right moving modes are no longer 
anti-parallel. At a given energy $E$ (not in the gap i.e.,~$|E|<|b_{_X}|$), 
we find the spin orientation $\vec S_P = \la~ \psi_{P} \,|\, \vec S \,|\,
 \psi_{P} ~\ra$ for the right/left movers ($P=R/L$) where $\vec S = \vec \si/2$
 is  the spin operator, to be given by  $S_{R/L}=(b_{_X},0,\pm\til k_0)/2E$.
When $|E| \gg |b_{_X}|$, the spin for the $R/L$ movers point  along the 
$Z/- Z$ direction as expected. 
Also,  note that the states at the bottom of upper~band and top of lower~band 
are located at $k =-b_{_Z}$ which implies that $\til k_0=0$ ; here, the respective 
spins point parallel and anti-parallel to the $X$-axis. For this value
 of the momentum,  the momentum dependent pseudo-magnetic field acting on the
 electron due to the helical  nature of the free Hamiltonian is fully canceled 
by the $Z$ component of the applied magnetic field; hence  it leaves  behind 
a net $\vec B$-field pointing along the $X$ direction~\cite{khari}. On a more 
general note, in a given band, if we look at left-moving and right-moving 
modes with the same energy, the magnetic field component $B_{_X}$ twists them by
the same angle, but  in opposite directions as sketched in Fig.~\ref{schematic}.
\begin{figure}[htb]
\includegraphics[scale=0.43]{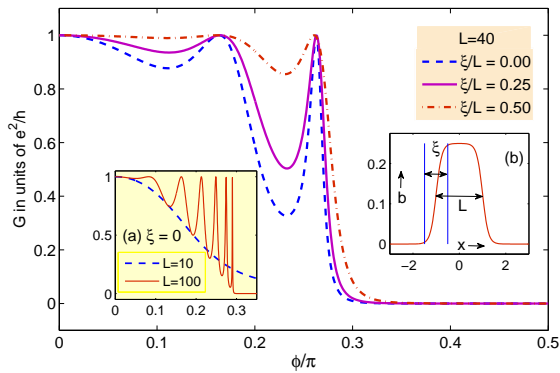}
\caption{ Main figure: Differential conductance $G$ vs angle $\phi$
for different $\xi/L$ . Inset(a): $G$ vs
$\phi$ for two different lengths of the patch with $\xi=0$. Inset(b): 
Illustration of how $b=|\vec b|$ changes across the patch for a fixed 
$\phi$. Parameters- $b_0=max.\{|\vec b|\}= 0.25$ and $E_i=0.2$ are the same 
in all the plots. }
\label{res}
\end{figure}

{\it{Transmission through the magnetic patch :-}} Let us consider the case 
where an electron is incident on the $\vec B$-field patch from the left with 
an energy $E_i$ and momentum $k_i=E_i$. Then the corresponding left and right
moving momentum eigenstates that  the incident particle excites in the patch 
region can be read off from Eq.~\eqref{LR-ud} with $E=E_i$.  The scattering 
states in different regions are given by 
\beq \psi = 
\begin{cases}  \vert \uparrow \, \ra \, e^{i{k_i}x} + r_{k_i} 
\vert \downarrow \,\ra \, e^{-i{k_i} x} &  \text{for $x<-L/2$,} \\ A_R \, 
\psi_{E_i\,R} \, e^{ik_R x} + A_L\, \psi_{E_i\,L} \, e^{-ik_L x} 
&\text{for $|x| < L/2$,}\\  t_{k_i} \, e^{i{k_i}x} \vert\uparrow \, \ra 
&\text{for $x > L/2$,} \end{cases}
 \label{LR} \eeq
where $\vert\uparrow\,\ra$ and $\vert\downarrow\,\ra$ are eigenstates
 of $\sigma_{_Z}$, $r$ and $t$ are the reflection and the transmission amplitudes
 and $A_L$ and $A_R$ are the  amplitudes corresponding to the left and right 
moving eigenstates in the $\vec B$-field patch. Using appropriate boundary
 conditions, one gets the following expression for transmission amplitude-
\bea t_{k_{i}} &=& \f{\til k e^{-i(k_i+b_{_Z})L}}{\til k \cos[\til k L]-ik_{i} 
\sin[\til k L]} ~.
\label{trans}
\eea
Note that the resonance  corresponds to 
$(k_R+k_L)L= 2 n \pi$ which is the total phase picked up by the electron
 in one round trip journey across the ${\vec B}$ field patch, $i.e.$, very 
similar to the double barrier resonance. The crucial difference between the 
two emerges when we look at the dynamics of the spin in the $\vec B$-field 
patch. This will be addressed below. The most interesting point to note here 
is the fact  that the resonance condition does not depend on $|\vec{B}|$ but 
only on $B_{_X}$. Hence the resonances can be tuned simply by rotating the 
${\vec B}$-field  away or toward the $Z$-axis without changing its magnitude.
This observation  is the central message of this letter.  A plot of the 
transmission probability at fixed $k_i$, which is  the differential 
conductance in units of $e^2/h$ evaluated at fixed bias, corresponding to 
an electron incident with energy $E_i$ from the left 
reservoir while the right reservoir is held at zero potential, is shown 
as a function of $\phi$ in Fig.~{\ref{res}}-inset-a. Here $\phi$ is the angle
 the $\vec B$-field makes with the $Z$-axis in the $X-Z$ plane. As we 
can see  from Fig.~\ref{res}-inset-a the conductance shows sharp resonances 
as we vary  $\phi$ for $L=100$. Note that the resonances get sharper
as $b_{_X}$ approaches $E_i$ from  $b_{_X}=0$, and beyond $E_i$, transport
 is subgapped. 
On the other hand note that for a short $\vec B$-field patch, ($L=10$ in 
Fig.~{\ref{res}}-inset-a), the transport is non-resonant, but the differential 
conductance can be as large as $0.1~e^2/h$ in the subgapped regime 
($\phi>0.3\pi$). Further, we have studied these resonances for a more realistic
$\vec B$-field patch wherein magnetic field changes from zero to maximum 
smoothly over a length-scale $\xi$ as illustrated in Fig.~\ref{res}-inset-b. 
We see  from Fig.~\ref{res} that the visibility of the resonances is affected
 by increasing $\xi$. However, even the least visible first resonance
(around $\phi=0.18\pi$), which is most affected by increasing $\xi$ is still visible
for $\xi$ in the range $(0,0.25)$.

{\it{Spin orientation in the magnetic patch :-}} 
In the presence of a magnetic field, the kinetic energy term competes with the 
Zeeman term and as a result, the left  and right movers point in different directions. 
(see Eq.~{\eqref{LR-ud}}). Hence when an electron is incident from the left on 
the $\vec B$-field patch, its spin  is expected to precess as it undergoes multiple
 reflections inside the magnetic  field patch due to the mismatch of the spin
 orientation of the incident electron and the spin orientations of the right and
left moving modes present inside the patch. But at resonance, the vanishing of
the reflection amplitude for the incident electron exactly amounts to an integer
number of complete spin precessions of the electron wave inside the magnetic field 
patch. Thus, at resonance,  the incident spin rotates back to its original 
orientation at the end of its journey inside the patch. We will show below that 
the precession dynamics in the $\vec B$-field patch is unconventional and carries
 nontrivial signatures of the spin anisotropy  of the HES.
\begin{figure}[htb]
\includegraphics[scale=0.30]{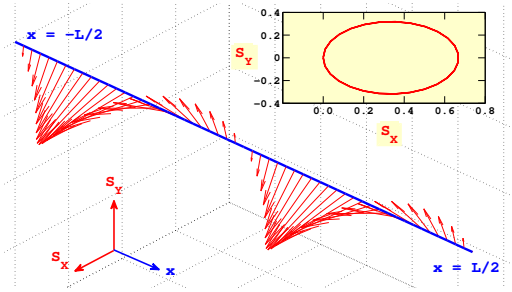}
\caption{ $X$ and $Y$ components of spin-density at different points in the 
magnetic-patch for  resonant transport $b_{_X}=0.1, b_{_Z}=0, L=180, k=0.0366$.
The inset shows $S_{_X}$ and $S_{_Y}$ at different values of  $x$, $S_{_Y}=0$ at
$x=-L/2,-L/4,0,L/4,L/2$ while $S_{_X}=0$ at $x=-L/2,0,L/2$.}
\label{sphelix}
\end{figure}
We first start by computing the spin orientation at each point $x$ between
 $x=-L/2$ and $x=L/2$ using the wave-functions given in Eq.~\eqref{LR}. Note
that since the wave function in the region of magnetic field is a linear
combination of left-moving and right-moving eigenstates, the spin at any given 
$x$ can, in general,  point out of the $X-Z$ plane even though the left and
right moving eigen-states are restricted to the $X-Z$ plane. Hence evaluating 
$\la~\psi\,|\,\vec \si/2\,|\,\psi~\ra$ 
for the $\vec B$-field patch using the wavefunctions given in Eq.~\eqref{LR} 
corresponding to an electron incident from the left with momentum $k_i$, we 
find 
\bea \la~ \vec S_{k_i}(x)~ \ra &=&\f {1}{2}  \f{1}{{\til k}^2
+b_{_X}^2\sin^2(\theta/2)}  \cdot  \nn \\
&& \!\!\!\!\!\!\!\! \big( b_{_X}k_i[1-\cos(\til \theta)], -2\,\til k\, b_{_X}  
\sin{\til\theta}, \til k^2 \big)~,  \label{sk}\eea
where $\til \theta = (x/L -1/2)\theta$ and $\theta=2\, \til k\,L$. It can be 
seen that as $x$ is changed from $-L/2$ to $L/2$, the $S_{_Z}$-component remains 
unaltered while the $S_{_X}$ and $S_{_Y}$ components trace out  the  locus of  
an ellipse.
Moreover, the spin density $\vec S_k(x)$ at any point $x$ makes a constant angle
with the axis $\vec n_k = (b_{_X},0,k)\cdot1/\sqrt{k^2+b_{_X}^2}$. Hence by
analogy to the case of precession of the spin of a static spin-half particle 
in a magnetic field, we see that   $\vec n_k$ defines the direction of  the
effective magnetic field around which  the spin density precesses  as a
function of $x$ as we move along the $\vec B$-field patch. Note that this 
precession is quite novel in the sense that the component of the spin  vector
along the precession axis $\vec n_k$ varies as the spin precesses, whereas
the component along the $Z$-axis is preserved. This is distinct from the
usual case where the component along the precession axis is preserved. 
The elliptical shape of the $X-Y$ projection of the precession
 is a mere reflection of the fact that the $Z$-axis is not the same as the 
precession axis. Finally, note that for the resonant case, $\theta=2n\pi$ at 
$x=\pm L/2$. This implies that $\til \theta$ is also $2 n \pi$, and hence,  
$\vec S_{k_i}=\hat Z/2$ (see Eq.~\eqref{sk}) at both $L/2$ and $-L/2$. 
This implies that the angle 
of precession of the electron in the resonant case is an exact integer
multiple of $2\pi$ which is very similar in spirit to the operational
principle of the Datta-Das transistor~\cite{DattaDas}. Note that even for any
finite transmission $\vec S_{k_i}$ must always point along $\hat Z$ for
$x\ge L/2$.  But if there is finite reflection at $x=-L/2$ then the spin for
$x\le-L/2$ definitely points away from  $\hat Z$ as the reflected wave 
always populates spin down states due to the 
helical nature of the edge states. So it is clear that for finite reflection, 
the direction of $\vec S_{k_i}$ at $x=-L/2$ can never be the same as that of
$x=L/2$. Thus, the resonance condition is directly coupled to evolution of spin
 density along the $\vec B$-field patch. This implies that our set-up can be 
regarded as a possible  candidate for devising a spin-transistor.\\
 {\it {Conditions for experimental observation:-}}
(i)~The ratio $\xi/L$ has to be small. $\xi/L<0.25 $ would be a reasonable 
limit for a typical case as mentioned earlier. (ii)~ The magnetic field that 
opens up a gap in the spectrum should not force the edge states to tunnel 
into the bulk. This means that the gap opened on the edge by $\vec B-$field 
should be less than bulk gap ($2|b_{_X}|<1$) and the energy of the incident
electron should also be within the bulk gap ($|E_i|<1/2$). A limit on the 
magnetic field then can be estimated to be 7 Tesla~\cite{yaku}. 
\begin{figure}
\includegraphics[scale=0.47]{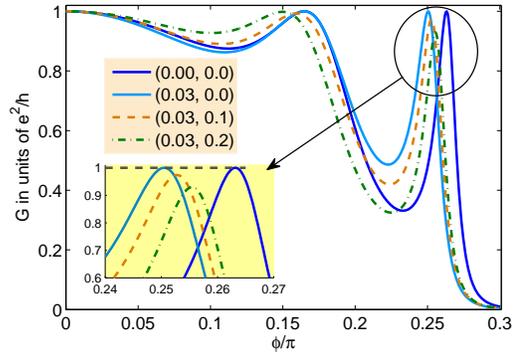}
\caption{Illustration of change in visibility of resonances 
due  to disorder of strength $\eta=0.03$ in the patch. 
Parameters chosen are same as that in Fig.~\ref{res} except 
$\xi/L=0.05,~l/L=0.1$. The legend shows $(\eta,x_l/L)$ for different 
curves.}
\label{diso}
\end{figure}
(iii)~Disorder:   The topological protection against 
back-scattering due to scalar impurities breaks down in the presence of
the time-reversal breaking magnetic field. Hence one cannot naively
throw away the disorder term. In a good sample we expect the disorder to 
be weak, sparsely spaced and to have a length-scale $l \ll L$. Such an 
impurity can be modelled by a rectangular potential-barrier/well with 
width $l$ and height $\eta$. This impurity may be centered anywhere 
($|x_l|<L/2$) in the magnetic field patch. We have studied the effect of 
such an impurity with a fixed $l$ with different disorder-strengths $\eta$,
 positioned across the patch ($|x_l|<L/2$). We find that as long as 
 $|\eta|\ll|E_i|$ the resonances are not affected. We have  given an 
 illustrative plot in Fig.~\ref{diso} of how the resonances 
 are affected for a value of $\eta/E_i=0.15$ positioned in the
  patch~\cite{us}. \\
{\it{Discussion and conclusion:-}}
The most crucial element involved in devising the proposed set up is to realize 
the localized $\vec B$-field patch whose direction should be fully tunable. It should
be possible to engineer such a localized  $\vec B$-field patch using the proximity 
effect~\cite{proximity} by depositing  a layer of magnetic material on top of the 
edge. Moreover,  by choosing a magnetic material which shows current induced rotation
 of magnetization ~\cite{domainwall}, it should be possible to  rotate the direction 
of the $\vec B$-field in the  patch. So in conclusion,  we have discussed a concrete 
proposal to probe the degree of spin anisotropy in the HES via the Fabry-P\'erot 
resonances. Note that  similar surface states appear in 3-D topological insulators
 and the extent to which such states are helical (spin-momentum locked) is not yet
 fully understood. Also, these have been probed not via direct transport 
experiments but via optics experiments such as spin-resolved ARPES which see 
a deviation\cite{arpes} from the purely theoretical picture of the 2-D helical 
states. Hence in the context of 2-D  topological insulators where 1-D HES 
appear on the boundary, studying these resonances could lead to crucial 
information and characterization of deviations from the theoretically predicted
 helical nature, if any. Also as an application, our set-up provides a possible
 design for a resonant spin transistor.

{\it Acknowledgments .-} We thank Diptiman Sen, Amir Yacoby and Yuval Oreg for
 stimulating discussions. A.S. thanks HRI and DU, for kind hospitality during
  the stay and CSIR, India  for financial support. S.R. thanks ICTS, Bangalore
  for kind hospitality during the completion of this work.

\end{document}